\documentclass[draft,jgrga]{agutex}
\usepackage{lineno}
\usepackage{graphicx}
\usepackage{multirow}
\usepackage{color} 
\usepackage{amssymb}
\usepackage{amsmath}

\authorrunninghead{Carlson et al.}
\titlerunninghead{PRELIMINARY BREAKDOWN SIMULATIONS}

\authoraddr{B. Carlson, Department of Physics, Carthage College, 2001 Alford Park Dr., Kenosha, WI, 53140, USA}
\authoraddr{C. Liang, STAR Lab, Electrical Engineering Department, Stanford University, Stanford, CA, USA}
\authoraddr{P. Bitzer, H. Christian, NSSTC, University of Alabama in Huntsville, Huntsville, AL, USA}

\begin{document}

\title{Time domain simulations of preliminary breakdown pulses in natural lightning}

\authors{B.E. Carlson, \altaffilmark{1,2}
C. Liang, \altaffilmark{3}
P. Bitzer, \altaffilmark{4}
and H. Christian \altaffilmark{4}} 

\altaffiltext{1}{Department of Physics, Carthage College, Kenosha, Wisconsin, USA}
\altaffiltext{2}{Birkeland Center for Space Science, University of Bergen, Bergen, Norway}
\altaffiltext{3}{Electrical Engineering Department, Stanford University, Stanford, California, USA}
\altaffiltext{4}{Department of Atmospheric Science, University of Alabama in Huntsville, Huntsville, Alabama, USA}
\altaffiltext{5}{Earth Systems Science Center, University of Alabama in Huntsville, Huntsville, Alabama, USA}

\begin{abstract}
  Lightning discharge is a complicated process with relevant physical scales spanning many orders of magnitude.  In an effort to understand the electrodynamics of lightning and connect physical properties of the channel to observed behavior, we construct a simulation of charge and current flow on a narrow conducting channel embedded in 3-dimensional space with the time-domain electric field integral equation, the method of moments, and the thin wire approximation.  The method includes approximate treatment of resistance evolution due to lightning channel heating and the corona sheath of charge surrounding the lightning channel.  Focusing our attention on preliminary breakdown in natural lightning by simulating step-wise channel extension with a simplified geometry, our simulation reproduces the broad features observed in data collected with the Huntsville Alabama Marx Meter Array.  Some deviations in pulse shape details are evident, suggesting future work focusing on the detailed properties of the stepping mechanism.
\end{abstract}

\begin{article}

\section{Introduction}
Despite centuries of study, many of the fundamentals of lightning physics are poorly-understood.  Much recent study has been devoted to measurements such as return stroke peak currents (e.g. \citet{Schoene2010}) or impulsive charge moment changes (e.g. \citet{cummer2004lcm}), and such descriptive results are extremely useful.  A comprehensive review including references to such descriptive results can be found in Chapters 4, 5, and 9 of \citet{Rakov2007}.  However, such descriptions do not necessarily help understand the fundamental physics of lightning.  The difficulty lies in capturing the fundamental physics, for example electron detachment and attachment rates, within a framework capable of reproducing lightning behavior.  Doing so involves a cascade of physical scales ranging from sub-millimeter-scale electron avalanches governed by micro-physical processes with measurable cross sections to 100 km-scale plasma channels governed by complex aggregate physical properties such as conductivity and charge density.  The general goal of this work is to describe and apply a large-scale simulation technique to connect the observed properties of lightning to physical properties of the channel that can be directly compared to existing understanding of the micro-physics.  The work here will not completely bridge the gap, but the techniques described herein can be broadly applied to many problems of lightning research.

The specific process considered in this paper is the step-wise extension of the lightning channel.  Though the exact mechanism is not known, the lightning leader channel tends to grow in length by impulsive steps whereby the channel lengthens suddenly, jumping forward into space ahead of the existing channel (see Chapter 4 of \citet{Rakov2007} for an overview).  Such steps range in length from tens to hundreds of meters and occur on timescales of order 10 $\mu$s.  One possible mechanism of such stepping is gradual heating of the gas near the channel tip by corona and streamer discharges.  This heating eventually reaches the point of thermally ionizing the gas, drastically increasing the conductivity.  An increase in conductivity results in an increase in current which acts to further heat the gas, further increasing the conductivity in a positive feedback effect, rapidly forming a new segment of conducting channel.  This development of the new segment of channel may occur slightly displaced from the end of the existing channel to form a disconnected channel dubbed the ``space stem.''  Recent observations include \citet{Hill2011}, who describe high-speed video observations of this stepping process and report the presence of such space stems ahead of the leader channel, while \citet{Winn2011} report balloon-borne electric field observations associated with lightning mapping data that are not completely consistent with such a connection process.  As such, at present the details of the stepping mechanism are not understood.  Once charges flow onto the narrow newly-heated channel, the charge density results in an outward electric force that drives excess charge outward to fill a ``corona sheath'' surrounding the channel.  This outward motion allows continued current flow onto the new segment, so the overall electrodynamics consist of a rapid increase in current as the new segment heats followed by a slower decrease in current as the corona sheath fills with charge.  These current and charge motions can be detected by electric field change meters as short pulses.

The goal of this paper is to understand and reproduce such pulses from preliminary breakdown in natural lightning as detected by the Huntsville Alabama Marx Meter Array (HAMMA) \citep{bitzer2013characterization}.  Preliminary breakdown here refers to electromagnetic emissions from natural lightning prior to the first return stroke.  These emissions are produced by a lightning channel growing within the cloud before it reaches the ground.  We approach this goal by self-consistently simulating the electrodynamics of the extension of an evolving channel with approximate treatment of channel plasma behavior.  Our work thus differs from most lightning modeling work, which most often either focuses on the return stroke where the channel is already highly-conducting or loses self-consistency by driving the channel with an assumed current source.  A review of return stroke literature can be found in \citet{baba2007emo}.  Existing stepped leader and preliminary breakdown models take a variety of approaches.  \citet{Karunarathne2014} approach preliminary breakdown pulses with a variety of modified transmission line models of preexisting channels driven by fitting the parameters of an assumed current source.  Note 20 of \citet{Baum1999} treats the leader step as a continuous extension at a given velocity within the framework of a segmented nonlinear transmission line model.  \citet{Gallimberti2002} characterizes each part of the system (leader, corona, space stem, etc.) and determines the time evolution of the characteristics as corona starts and stops and as the leader extends.  \citet{Kumar2000} describes the leader as a quasi-static system of charges and currents exhibiting RLC circuit behavior.  \citet{Larigaldie1992} considers the time domain electrodynamics and uses a similar treatment as in our approach, but focuses on strikes involving aircraft.  Our approach also distinctly differs from the quasi-static models of \citet{Niemeyer1984,Mansell2005,Riousset2007,krehbiel2008ued}, which disregard time evolution and therefore cannot determine current pulse shapes, electromagnetic wave emissions, or the time structure of channel development.

Our simulation, described in Section~\ref{sect:simulations}, is based directly on Maxwell's equations, explicitly includes time evolution of electric charge and current, and retains the dynamics of channel heating and charge migration away from the channel.  We approximate the details of channel behavior to limit the model complexity and the number of free parameters.  The remaining parameters are all physical properties of lightning channel behavior such as specific heat or the timescale for charge migration to the corona sheath.  The dependence of these results on the parameters of the simulation is described in Section~\ref{sect:params}.  We then use our simulation to predict possible electromagnetic emissions from stepwise channel extension in preliminary breakdown as described in Section~\ref{sect:results}.  The simulation results are compared with observations in Section~\ref{sect:comparison}.  The results of the comparison are used to suggest processes not properly captured in the simulation.  We then suggest future studies and conclude in Section~\ref{sect:discussion}.

\section{Simulations}
\label{sect:simulations}
The simulation technique described in this paper, which we dub time-domain fractal lightning (TDFL) modelling, is an electrodynamic simulation of charge and current flow on a narrow branched conducting channel embedded in 3-dimensional space capable of reproducing fractal lightning geometry.  (For an alternative approach to TDFL modelling as used to describe return strokes, see \citet{Liang2014}.)  The simulation acts on the assumption that the electric field is the dominant driver of the electrodynamic behavior of lightning.  This assumption is justified by the weakness of magnetic forces relative to electric forces, especially when charge imbalances are present as is the case for lightning.  The simulation is described in detail as follows, but overall proceeds as a series of time-steps.  In each step, the history of the channel is used to determine the electric field.  The electric field is then used to determine the current evolution during the time-step.  The current then determines how charge distributions evolve during the time-step.  The charges and currents thus determined are then recorded, and the process is repeated to determine the time evolution of the system.  This scheme is implemented as a method of moments solution to the electric field integral equation with the thin wire approximation and marching on in time, largely following and extending the method described in \citet{Miller1973}.

More specifically, in general electric fields can be calculated using the retarded-time electric field integral equation (EFIE, equation 6.55 in \citet{Jackson1999}):
\begin{equation}
\mathbf{E_\mathrm{t}}(\mathbf{x},t) = \frac{1}{4 \pi \epsilon_0} \int d^3 x' \left \{ \frac{\mathbf{\hat{R}}}{R^2}\left[ \rho(\mathbf{x'},t') \right]_{\mathrm{ret}} + \frac{\mathbf{\hat{R}}}{cR} \left[\frac{\partial \rho(\mathbf{x'},t')}{\partial t'} \right]_{\mathrm{ret}} - \frac{1}{c^2 R} \left[\frac{\partial \mathbf{J}(\mathbf{x'},t')}{\partial t'} \right]_{\mathrm{ret}} \right \}
\end{equation}
where $\mathbf{E_\mathrm{t}}$ is the total electric field, $\mathbf{x}$ is the observation point, $t$ is the observation time, $\mathbf{x'}$ is the source point, $d^3x'$ signifies integration over sources at points in (three dimensional) space, $\mathbf{R} = \mathbf{x}-\mathbf{x'}$ is a vector from source to observation location, $R = |\mathbf{R}|$, $\mathbf{\hat{R}} = \mathbf{R}/R$ is the corresponding unit vector, $t' = t - R/c$ is the (retarded) time at the source, $\epsilon_0$ is the permittivity of free space, $c$ is the speed of light, $\rho$ is charge density, $\mathbf{J}$ is current density, and $[...]_\mathbf{ret}$ emphasizes the evaluation at retarded time $t'$.  The EFIE is a Green's function solution to the full set of Maxwell's equations, where the first term $\propto 1/R^2$ is the familiar static electric field while the last two terms derive from time derivatives of magnetic fields and contribute inductive and radiative effects.  The use of a full time-domain solution to Maxwell's equations allows us to treat current and charge on the lightning channel as functions of time, in contrast to solution techniques based on the Poisson equation.

The EFIE must be integrated over all space, but since we wish to focus our attention on lightning, we treat charges not directly associated with the lightning channel as an external static applied electric field, $\mathbf{E_\mathrm{a}}$, which adds to the lightning electric field $\mathbf{E_\mathrm{l}}$ to give the total electric field $\mathbf{E_\mathrm{t}}$.  $\mathbf{E_\mathrm{a}}$ is treated as an input to the simulation that encompasses the static effects of thunderstorm charge centers, screening charge layers, and net charges in the ground beneath the storm.  Since in this work we consider channels and timescales that are short compared to the scale of variability of cloud charge density, $\mathbf{E_\mathrm{a}}$ is taken to be constant.

The electric field $\mathbf{E_\mathrm{l}}$ associated with the lightning channel is where we must apply the EFIE, observing that the lightning channel is effectively a long narrow structure.  Partially integrating the EFIE over the cross sectional area of the channel converts the three dimensional volume integral into a one-dimensional integral along the channel:
\begin{equation}
\mathbf{E_\mathrm{l}}(\mathbf{x},t) = \frac{1}{4 \pi \epsilon_0} \int ds' \left \{ \frac{\mathbf{\hat{R}}}{R^2}\left[ \lambda(s',t') \right]_{\mathrm{ret}} + \frac{\mathbf{\hat{R}}}{cR} \left[\frac{\partial \lambda(s',t')}{\partial t'} \right]_{\mathrm{ret}} - \frac{1}{c^2 R} \left[\frac{\partial \mathbf{I}(s',t')}{\partial t'} \right]_{\mathrm{ret}} \right \}
\label{eqn:efie1d}
\end{equation}
where volume charge density $\rho$ has become linear charge density $\lambda$, vector current density $\mathbf{J}$ has become a vector total current $\mathbf{I}$ directed along the channel, and the three degrees of freedom of the vector $\mathbf{x'}$ are now represented as a scalar length coordinate $s'$ specifying position along the channel.

\begin{figure*}
\begin{center}
\includegraphics[width=6.0in]{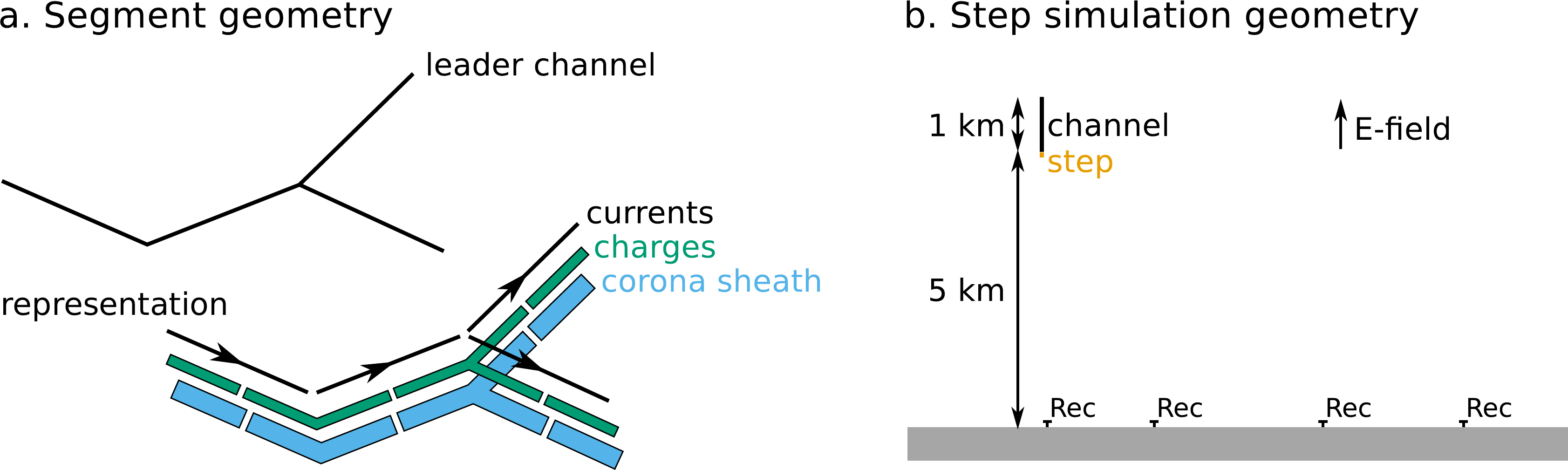}
\end{center}
\caption{(a) Channel discretization and representation geometry.  The lightning channel is divided into straight current segments that connect charges, where charges are groups of straight segments and are surrounded by corona sheath segments with the same basic geometry as the associated charge segments. (b) Simulation geometry.  The simulations in this paper largely consider a straight 1 km vertical channel placed 5 km above a perfectly conducting ground with receivers (``Rec'' in the figure) placed at ground level to record the vertical electric field.}
\label{fig:seggeom}
\end{figure*}

This one-dimensional integral version of the EFIE is unfortunately still not mathematically tractable, so we treat the EFIE numerically with the method of moments \citep{rao1999time} by dividing the channel into current and charge segments.  Each current segment flows into a charge and out of a charge.  Charges are represented as groups of straight segments partially overlapping the connected current segments as shown in Figure~\ref{fig:seggeom}a.  Current and charge density are assumed to be uniform over the charge and current segments and thus piecewise uniform over the channel.  This assumption means that the 1-d EFIE can be further simplified, separating the integral into a sum of many relatively simple sub-integrals, one over each charge or current segment.  For example, consider the electrostatic term in the EFIE:
\begin{equation}
  \frac{1}{4 \pi \epsilon_0} \int ds' \frac{\mathbf{\hat{R}}}{R^2}\left[ \lambda(s',t') \right]_{\mathrm{ret}} = \frac{1}{4\pi\epsilon_0} \sum_i \int_i ds' \frac{\mathbf{\hat{R}}}{R^2} \left[ \lambda_i(t') \right]_{\mathrm{ret}}
\end{equation}
where $i$ is an index for the set of charge segments, $\int_i$ represents integration over segment $i$, and $\lambda_i(t')$ is the charge density of segment $i$, which is uniform and thus only a function of $s'$ through the effect of $s'$ on the retarded time $t'$.  We further simplify this by assuming the segment is short compared to the speed of light timescale of the processes to be captured by the simulation, so the effects of retarded time do not vary significantly over a segment.  This means the retarded time charge density can be factored out of the integral:
\begin{equation}
  \frac{1}{4\pi\epsilon_0} \sum_i  \left[ \lambda_i(t') \right]_{\mathrm{ret}} \int_i ds' \frac{\mathbf{\hat{R}}}{R^2}
\end{equation}
leaving the integral solely to treat geometric effects.  Applying this process to the $\partial\lambda/\partial t'$ and $\partial\mathbf{I}/\partial t'$ terms, the EFIE in full is:
\begin{equation}
  \mathbf{E_\mathrm{l}}(\mathbf{x},t) = \frac{1}{4\pi\epsilon_0}\left\{
    \sum_i\left(\left[ \lambda_i(t')\right]_\mathrm{ret} \int_i ds' \frac{\mathbf{\hat{R}}}{R^2}\right)
    + \sum_i\left(\left[ \frac{\partial \lambda_i(t')}{\partial t'}\right]_\mathrm{ret} \int_i ds' \frac{\mathbf{\hat{R}}}{cR}\right)
    - \sum_i\left(\left[ \frac{\partial I_i(t')}{\partial t'}\right]_\mathrm{ret}\mathbf{\hat{I_i}} \int_i \frac{ds'}{c^2R}\right)
  \right\}
\end{equation}
where $\mathbf{\hat{I_i}}$ is a unit vector capturing the direction of current segment $i$.  Note that a system with $N$ current segments has $N+1$ charges, so the first two sums over $i$ run up to $N+1$ while the last sum only runs up to $N$.  Assuming the lightning channel does not physically move significantly during a discharge, the geometric factors as calculated by the integrals above do not evolve with time.  As such, they can be calculated once and treated simply as constants in the simulation, leaving the resulting equation simply a sum of geometric factors (constants) multiplied by charges and currents, a linear equation.

\begin{figure}
\includegraphics[width=3.5in]{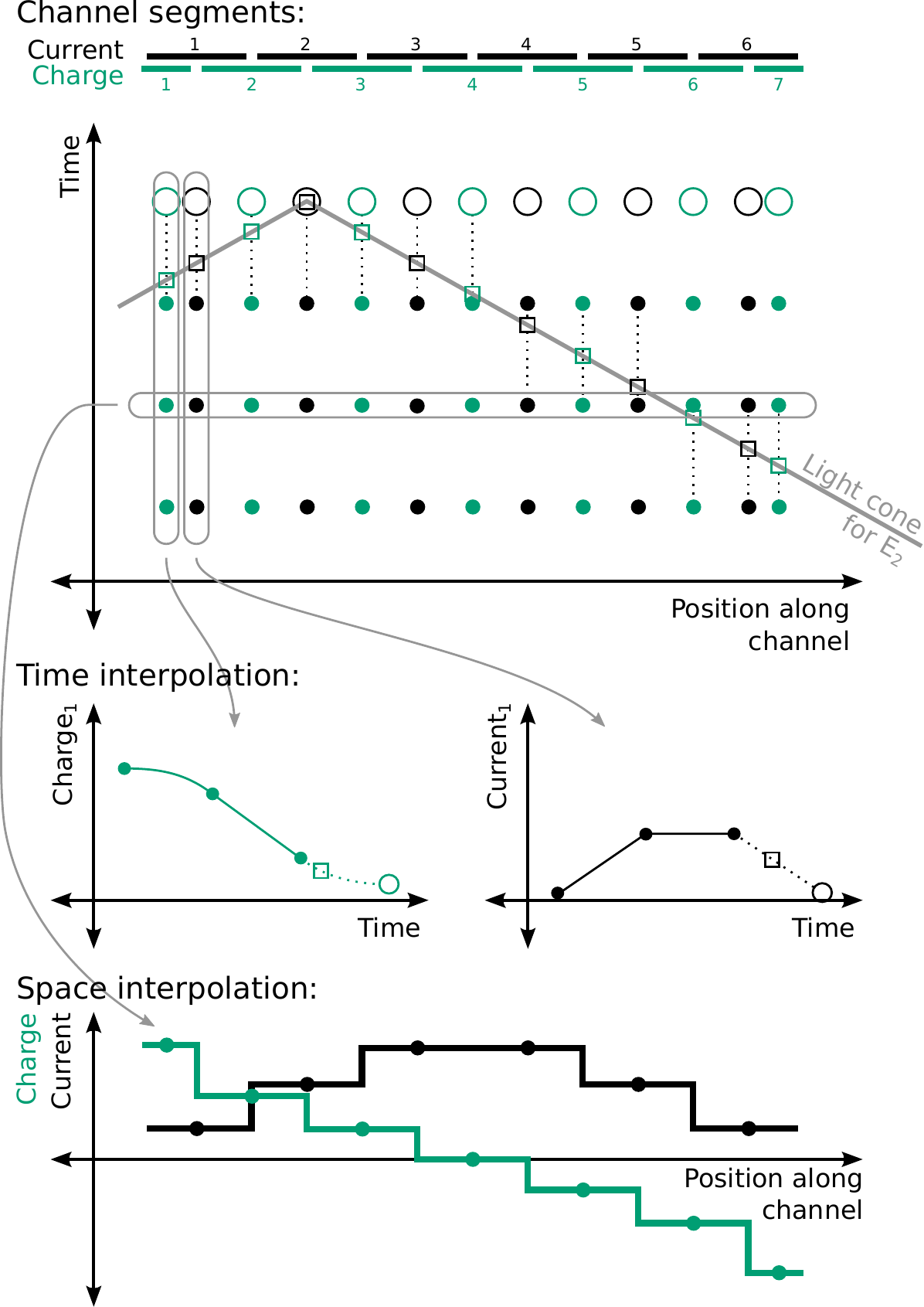}
\caption{Time and space discretization, interpolation, and retarded time integration.  The current and charge segment geometry is shown at the top, aligned with the space/time grid points where charge and current values are recorded shown below (filled green and black circles represent known history of charge and current respectively, open green and black circles represent unknown charge and current values, respectively, that are needed for time evolution).  An example EFIE retarded time integral to calculate the electric field at the center of current segment 2 is shown in grey (labeled ``light cone''), as are the points where interpolated charge and current values are used (open squares).  Slices through this space/time grid to demonstrate the interpolation scheme used are shown at the bottom.}
\label{fig:timing}
\end{figure}

The time evolution of the system is also discretized: currents and charges are recorded at a set of times $t_j$ separated by a time-step $\delta t$ at locations represented by the center point of the segment in question, see Figure~\ref{fig:timing}.  Time evolution of channel current is assumed to be piecewise linear (piecewise constant current time derivative), leading to quadratic time-variation in charge density.  The interpolation scheme is implemented as a set of basis functions (e.g. triangle functions for linear interpolation) whereby the desired quantity at a desired time (one of the filled squares in Figure~\ref{fig:timing}) is determined as a sum of the values in the grid before and after the point in question (i.e. those points connected by a dotted line with the filled square in question) multiplied by the appropriate interpolation basis function.  Such interpolation basis functions can capture the time derivative behavior as well, and the resulting scheme is purely linear, meaning the contribution to the electric field of each point in the space/time grid is simply proportional to the value at that grid point.

The assumption that the lightning channel does not move significantly during a discharge also means the interpolation basis function evaluations necessary to apply the EFIE can also be effectively represented as a set of constants, fully reducing the EFIE to its form used here:

\begin{equation}
  \mathbf{E_\mathrm{l}}(\mathbf{x},t_n) = \frac{1}{4\pi\epsilon_0}\left\{ \sum_i \left[ \left(\sum_j^n \lambda_i^j \alpha_i^j \mathbf{G}_{r^2}^i\right) + \left(\sum_j^n \lambda_i^j \beta_i^j \mathbf{G}_{rc}^i\right)\right] - \sum_i \left[\sum_j^n I_i^j \gamma_i^j \mathbf{G}_{rc^2}^i\right]\right\}
\end{equation}

Where the first and second sums over $i$ replace the integral over the charges and currents in the channel respectively, the sum over $j$ carries out the sum over multiple grid times necessary to account for time interpolation to the desired non-grid times needed in the retarded time integral, $\mathbf{G}_{r^2}^i$, $\mathbf{G}_{rc}^i$, and $\mathbf{G}_{rc^2}^i$ are the geometric factors resulting from the spatial integral over segment $i$ as seen by an observer at point $\mathbf{x}$ (the subscript represents the denominator of the corresponding term in equation \ref{eqn:efie1d}), while $\alpha_i^j$, $\beta_i^j$, and $\gamma_i^j$ are the interpolation basis function evaluations necessary to find the contribution from charge, charge time derivative, and current time derivative respectively for segment $i$ at time-step $j$ as seen by an observer at point $\mathbf{x}$ at time $t_n$.  Note that due to the retarded time structure of the system, the $\alpha_i^j$, $\beta_i^j$, $\gamma_i^j$ factors are dependent on the position of the observation point so the above expression is specific to the value of $\mathbf{x}$, and also that most of the $\alpha_i^j$, $\beta_i^j$, $\gamma_i^j$ are zero, so in practice the computational complexity of this expression is that of a single sum instead of two nested sums.

The two additional ingredients needed to complete the simulation are Ohm's law and a method to determine the time evolution of channel current.  Ohm's law, $\mathbf{J} = \sigma\mathbf{E}_t$, relates current density and total electric field given a conductivity $\sigma$.  Taken over the cross sectional area of a narrow channel assuming current flows along the channel only and the electric field is approximately constant across the cross section, Ohm's law becomes $I = \sigma A \mathbf{E_\mathrm{t}}\cdot\mathbf{\hat{s}}$ giving the scalar current at the point in question in terms of the dot product of the total electric field, the cross sectional area $A$, and $\mathbf{\hat{s}}$, a unit vector giving the direction of the channel at the point in question.  Note that the resistance per unit length $R_l = 1/(\sigma A)$.

This suggests a method to determine currents in a given segment, supposing the first $n-1$ time-steps of the simulation are complete.  Apply the EFIE to determine the component of the electric field along the channel at the location of the given segment (segment $k$) for time-step $n$ and apply Ohm's law to determine the current.  This overall gives a linear equation:

\begin{equation}
I_k^n R_l = \frac{1}{4\pi\epsilon_0} \left\{\sum_i \left[ \left(\sum_j^n \lambda_i^j \alpha_{ik}^j G_{r^2\,k}^i\right) + \left(\sum_j^n \lambda_i^j \beta_{ik}^j G_{rc\,k}^i\right)\right] - \sum_i \left[\sum_j^n I_i^j \gamma_{ik}^j G_{rc^2\,k}^i\right]\right\}
  \label{eqn:efieReduced}
\end{equation}

Where the subscript $k$ indexes the segment of interest and the $G_{\cdots}^{\cdots}$ are now scalar contributions to the electric field component along segment $k$ instead of vectors giving the full electric field.  Note first that most of the terms on the right hand side are contributions from the charge and current in the past to the electric field in the present.  However, the sum over $j$ representing the sum over history runs all the way up to and includes time-step $n$, i.e. including influences from the recent past, nearby segments, and the contribution of a segment to itself.  This implies that the unknown $I_k^n$ appears on both the left- and right-hand sides of the equation (as in Figure~\ref{fig:timing} the integration for $E_2$ required $I_2^n$ for interpolation), and that the equation for $I_k^n$ will also involve nearby unknown currents (as seen in Figure~\ref{fig:timing} the integration for $E_2$ explicitly required $I_1$ and $I_3$ for interpolation).  Note finally that the contributions required from interpolations involving unknown charges can be expressed in terms of contributions from known past charges interpolated up to time-step $n$ by use of the unknown currents, adding further contributions from various unknown $I_i^n$.  These complications pose no serious problems, and in a system with $N$ current segments ($N+1$ charge segments), repeating this process for all current segments provides a system of $N$ linear equations in $N$ unknown currents that can easily be rearranged into a form convenient for numerical solution.

One final subtlety is the influence of a segment on itself.  Since the channel is effectively one-dimensional, the $1/R$ and $1/R^2$ terms in the geometric factors $G_{\cdots\,k}^k$ described above will diverge.  We solve this problem by making the thin wire approximation, $R \rightarrow \sqrt{R^2 + r_0^2}$ in the denominators of terms of the EFIE, where $r_0$ is the effective radius of the channel \citep{Miller1973}.  This approximation is valid if the typical value of $R$ is much greater than that of $r_0$, which it is even for such self-contributions if the segment length is much larger than the channel radius.  Since the typical radius of a lightning channel is a few mm \citep{Rakov1998} and segments used to represent a lightning channel are at least several meters long this approximation is clearly justified.  We further improve the calculation of geometric factors $G_{\cdots\,k}^i$ (i.e. those needed to calculate the electric field at segment $k$ due to segment $i$) by calculating the result of the integral over segment $i$ as observed by many points on segment $k$ and computing an average.  In our implementation, this averaging is especially important for proper calculation of segment self-contributions (i.e. $G_{\cdots\,k}^k$) which are very important for correct time-evolution features such as current wave speed.

The system of linear equations that results can easily be solved on the computer by matrix techniques.  The solution gives the currents at the next time-step.  The charge values at the next time-step can be determined by applying charge conservation along the channel and integrating the current flow into and/or out of a given segment over the given time-step in question.  The simulation keeps track of this by working internally with charges $Q_i$ instead of charge densities $\lambda_i$ ($Q_i = \lambda_i L_i$ where $L_i$ is the total length of charge segment $i$) and by defining a connectivity matrix $C_{il}$ defined as
\begin{equation}
  C_{il} = 
  \left\{
  \begin{array}{lr}
    +1 & \text{if } I_l \text{ flows into charge } Q_i \\
    -1 & \text{if } I_l \text{ flows out of charge } Q_i \\
    0 & \text{otherwise}
  \end{array}
  \right.
\end{equation}
Though such a matrix is unnecessarily complex for the simple unbranched geometry considered in this work (here $C_{i\ i}=-1$ and $C_{i+1\ i}=+1$ for $i=1\ldots N$), more complicated network connectivity can easily be captured.  It is also quite computationally convenient, for example giving the net charge flow into a segment as a matrix-vector product:
\begin{equation}
  \frac{dQ_i}{dt} = \sum_l C_{il}I_l
\end{equation}

With this definition of the connectivity matrix, charge conservation and integration forward for over a time-step is straightforward.  The piece-wise linear time interpolation for current integrated over a time-step simply becomes an average, giving
\begin{equation}
  Q_i^n = Q_i^{n-1} + \frac{1}{2}\left(\sum_l C_{il}I_l^n + \frac{1}{2}\sum_l C_{il}I_l^{n-1}\right)\delta t
\end{equation}
where the first sum gives the net current flow into charge segment $i$ at time $t_n$ and the second gives the same quantity at time $t_{n-1}$.  This expression gives time evolution for a simple case that will become more complicated when we add features to the model in the next section.  Regardless, these new current and charge values complete a new state of the system to be added to the history.  Repeating this procedure marches the system forward in time.

One further advantage of this system is that the time-step size is flexible.  For periods during long duration simulations when no short-duration processes are happening, the time-step can be extended, trading time resolution for simulation speed.  The system smoothly transitions from a full electrodynamic to a quasi-static simulation.  As time-step size increases, more and more nearby unknowns appear in equation \ref{eqn:efieReduced}, and when the time-step is infinitely large, the system of linear equations that results is equivalent to the solving the Poisson equation over the lightning channel.

We have verified our simulation as described above against time-domain results from \citet{Poggio1973} and static results from \citet{jackson2000cdo} with very good agreement.  While such time domain method of moments calculations often have high-frequency stability problems, we follow the time averaging scheme described in \citet{smith1990instabilities} to dampen out these high-frequency oscillations.  Von Neumann stability analysis of the method shows good stability characteristics provided the time-step is not too short, and stability improves for longer time-steps though resolution is lost.

The simulation itself is written in C with sparse matrix operations from CXSparse \citep{Davis2006} and is written to run in parallel on multi-CPU computers with ScaLAPACK \citep{slug} and MPI.

\subsection{Additional features for leader extension simulation}
The simulation system described above cannot immediately duplicate the features seen in natural lightning step pulses.  In particular, the corona sheath must be included in order to reproduce the quantity of charge transfered by a channel.  Capturing the details of the radial distribution of charge would vastly increase the computational complexity of the model, so here we simply include the corona sheath as a secondary set of charges, located on top of the main channel charges that enter the EFIE with the same formalism as described above but with a larger effective radius $r_\mathrm{CS}$ in the thin wire approximation as used in computation of geometric factors $G_{\cdots}^{\cdots}$.  It is important to note that this radius parameter is not the exact radius of a cylindrical sheath of charge, especially since this radius may become comparable to the segment length and thus leaving the thin wire approximation unjustified, but the corona sheath radius parameter does capture the behavior of a diffuse charge region surrounding the channel with effective size tunable by $r_\mathrm{CS}$.

A corona sheath charge segment is filled with charge from the channel charge segment it encloses, i.e. we assume charge flows only outward or inward from a given channel charge segment to its corona sheath, not longitudinally along the corona sheath.  Higher charge density on the main channel charge segment would lead to more rapid transfer, so we further assume the charge migrates outward at a rate proportional to the charge stored on the channel.  This effectively means we treat the corona discharge processes surrounding the channel as represented by a constant conductivity; this is not a good assumption but it is convenient: the linearity of the charge transfer process means it can be interpolated in time with the same sort of interpolation basis functions described above.  Ignoring charge flow along the channel, the overall result of this approach would be an exponential decrease in channel charge, so we parametrize this process by the characteristic timescale $\tau_\mathrm{CS}$ for charge transfer to the corona sheath.  Thus, the time derivative of the charge on corona sheath segment $i$ ($Q_{i\mathrm{CS}}$) is given by
\begin{equation}
  \frac{dQ_{i\mathrm{CS}}}{dt} = \frac{Q_i}{\tau_\mathrm{CS}}
\end{equation}
where $Q_i$ is the charge on the corresponding channel segment.  The charges in the equation above are functions of time, so time evolution of the system becomes more complicated than described above.  Instead of simply integrating the current flow along the channel into or out of a channel charge segment to determine the change in the charge, we must solve a differential equation for the evolution from one timestep ($t_{n-1}$) to the next ($t_n$):
\begin{equation}
  \frac{dQ_i}{dt} = -\frac{Q_i}{\tau_\mathrm{CS}} + \mathcal{I}_i^{n-1}\left[1-\frac{t-t_{n-1}}{\delta t}\right] - \mathcal{I}_i^n \left[\frac{t-t_{n-1}}{\delta t}\right]
\end{equation}
where $-\frac{Q_i}{\tau_\mathrm{CS}}$ gives the current flow outward to the corona sheath due to the charge on the segment in question, $\mathcal{I}_i^{n-1} = \sum_l C_{il}I_l^{n-1}$ and $\mathcal{I}_i^n = \sum_l C_{il}I_l^n$ are the net current flow into the charge segment at times $t_n$ and $t_{n-1}$, respectively, in terms of the connectivity matrix $C_{il}$ as discussed previously, and the terms in square brackets are the piecewise linear time interpolation basis functions discussed above appropriate for interpolation between $I_i^{n-1}$ and $I_i^n$ for times between $t_{n-1}$ and $t_n$.  The equation is a first-order non-homogeneous ordinary differential equation that is straightforward to solve by variation of parameters.  The result, valid for times between $t_{n-1}$ and $t_n$, is
\begin{equation}
  Q_i(t) = \left(Q_i^{n-1} - \mathcal{I}_i^{n-1} \tau_\mathrm{CS} + \frac{\mathcal{I}_i^n - \mathcal{I}_i^{n-1}}{\delta t}\tau_\mathrm{CS}^2\right)e^{-\frac{t-t_{n-1}}{\tau_\mathrm{CS}}} + \mathcal{I}_i^{n-1} \tau_\mathrm{CS} + \frac{\mathcal{I}_i^n - \mathcal{I}_i^{n-1}}{\delta t}\left((t-t_{n-1})\tau_\mathrm{CS} - \tau_\mathrm{CS}^2\right)
\end{equation}
Evaluating this solution at $t_n = t_{n-1} + \delta t$ thus determines the evolution of the net charge carried on segment $i$ from time $t_{n-1}$ to time $t_n$.  The evolution of charge on the corresponding corona sheath segment can then be computed easily by considering charge conservation on the channel segment in question:
\begin{equation}
  Q_{i\mathrm{CS}}^n = Q_{i\mathrm{CS}}^{n-1} + \left(\frac{\mathcal{I}_i^{n-1} + \mathcal{I}_i^n}2\delta t - (Q_i^n - Q_i^{n-1})\right)
\end{equation}
where $\frac{\mathcal{I}_i^{n-1} + \mathcal{I}_i^n}2\delta t$ is the net charge flow along the channel onto the channel charge segment in question during the time-step in question and $Q_i^n - Q_i^{n-1}$ is the net change in charge on the channel charge segment in question.  Any imbalance between these terms is due to charge flow onto the corona sheath.

Furthermore, in channel extension simulations, the channel itself must evolve with time.  This evolution is determined by heating and cooling processes that also must be included in the simulation in order to reproduce the features of step pulses.  Fundamentally, the heating process is Joule heating, with a power per length proportional to $I^2 R_l$, where $R_l$ is the resistance of the channel per unit length.  Cooling is determined by a combination of radiative, conductive, and convective cooling.  The fundamental physics of heating and cooling of a non-equilibrium plasma channel is very complex, and we have not attempted to capture its nuances here.  For a more detailed consideration, see \citet{Liang2014}.  Here we simply assume that the conductivity of the channel can be determined by an effective temperature, and that the effective temperature changes according to a effective heat capacity per unit length $C_l$.  On timescales shorter than 20~$\mu$s as examined here, cooling is not a major factor \citep[Chapter 6]{Heckman1992}, so we simply have $C_l dT/dt = I^2 R_l$.  In the segmented representation of the channel, a temperature is assigned to each current segment and evolved according to the heating at the end of each time-step.

\begin{figure}
\includegraphics[width=3.0in]{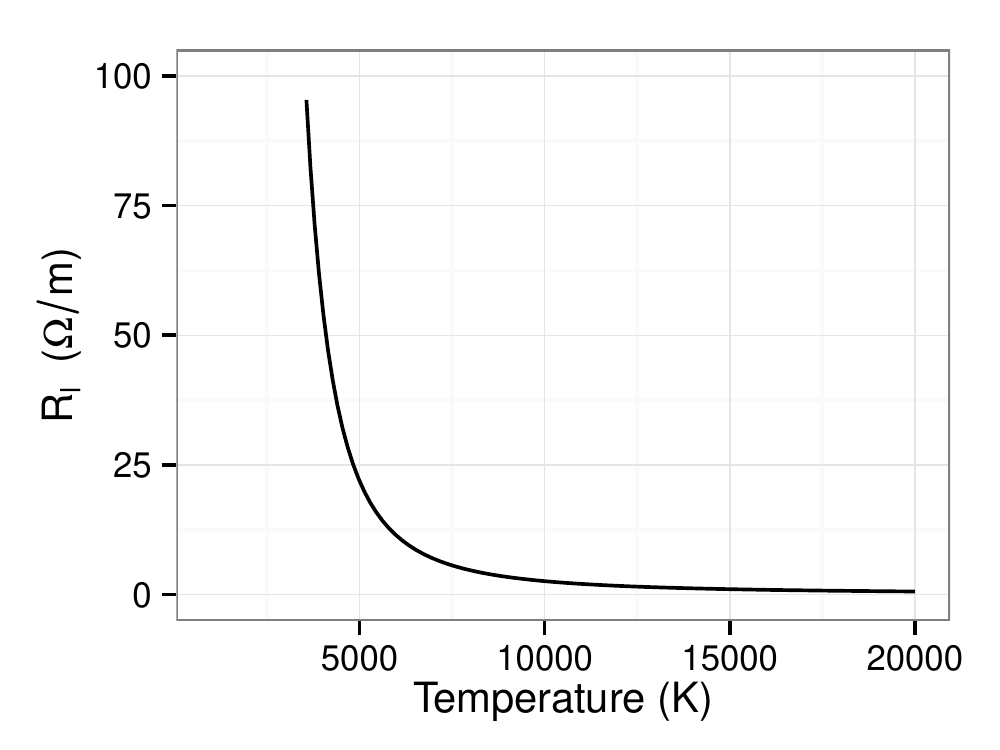}
\caption{The formulation of resistance per unit length of the channel as a function of temperature of the channel used in the simulations.}
\label{fig:rPerL}
\end{figure}

Given a temperature, it remains to calculate the resistance per unit length of the channel, $R_l$.  This is another complex topic that we can only qualitatively approximate.  The conductivity calculation here is motivated by the Saha equation of ionization equilibrium and results from plasma conductivity studies.  The Saha equation gives ratios of various ionization states in terms of in terms of their degeneracies and thermal energy effects, but if the temperature dependence is the only effect of importance, it becomes simply a proportionality $n_e^2 \propto T^{3/2}e^{-\epsilon/k_\mathrm{B} T}$ where $n_e$ is the electron number density, $T$ is the temperature, $k_\mathrm{B}$ is Boltzmann's constant, and $\epsilon \approx 14$~eV is the approximate ionization energy relevant for atomic oxygen or nitrogen.  The plasma conductivity adapted to non-ideal plasma conditions as in \citet{Zollweg1987} is typically given in terms of a reduced conductivity, $\sigma^{*}$ as $\sigma \propto T^{3/2} \sigma^{*}$.  The reduced conductivity $\sigma^{*}$ is roughly proportional to the square root of a non-ideality parameter, $\sigma^{*} \propto \gamma^{1/2}$ (see Figure~1, \citet{Zollweg1987}), where $\gamma \propto n_e^{1/3}/T$.  Combining these proportionalities, we obtain $\sigma \propto T^{9/8} e^{-\epsilon/12 k_\mathrm{B}T}$ and a corresponding resistance per unit length $R_l \propto T^{-9/8} e^{\epsilon/12 k_\mathrm{B}T}$.  Comparison to arc conductivity measurements in \citet{Schreiber1973} adjusted to our channel radius give $R_l(10^4 \mathrm{K}) \approx 2.5$ $\Omega$/m sets the proportionality constant.  This calculation gives the resistance vs temperature shown in Figure~\ref{fig:rPerL}.  This is of course at best an approximate treatment, and changes in the resistance as a function of temperature may make channels heat up more or less quickly which will have an effect on the electromagnetic radiation produced.  Regardless, since channel heating is a relatively slow process compared to our time-step size, in each time-step, the current and charge are updated assuming constant temperature and conductivity.  The temperature and conductivity are then updated before the next time-step.

Though approximate, these treatments of corona sheath behavior and temperature-dependence of resistance capture the dominant behavior of the lightning channel.

\subsection{Parameters}
\label{sect:params}

Such a complex simulation naturally has parameters.  These parameters break down into 3 main categories: physical constants, computational parameters, and initial conditions.  The physical constants include not only true physical constants like the speed of light or the permittivity of free space, but also the physical properties of the channel that do not evolve with the simulation: channel heat capacity and effective radius for example.  The initial conditions (e.g. the initial geometry of the channel or the length of the step in question) are discussed later in the context of specific simulations.  A full list of the parameters and their values is given in Table~\ref{table:params0}.

\begin{table}
\caption{List of simulation parameters}
\label{table:params0}
\centering
\begin{tabular}{r|l}
  \multicolumn{2}{l}{Physical constants} \\ \hline
  Channel radius & 3 mm \\
  Corona sheath radius & 4 m \\
  Corona sheath timescale & 0.5 $\mu$s \\
  Channel heat capacity & 2 J/K/m \\
  & \\
  \multicolumn{2}{l}{Computational parameters} \\ \hline
  time-step size & 35 ns \\
  Segment length & 5 m \\
  & \\
  \multicolumn{2}{l}{Initial conditions} \\ \hline
  Applied electric field & $\sim 100$ kV/m \\
  Channel length & $\sim 1$ km \\
  Channel shape & straight \\
  Channel orientation & vertical or angled \\
  Channel position & above origin\\
  Step length & $\sim 100$ m \\
\end{tabular}
\end{table}

The channel radius is taken to be $3$ mm, consistent with \citet{Rakov1998}, though this parameter does not significantly affect the results.  The corona sheath radius (i.e. $r_\mathrm{CS}$ in the thin wire approximation) is taken to be $4$ m, consistent estimates from \citet[page 292]{cooray2004lf} based on the distance from the channel for electric fields to decrease below breakdown for typical linear charge densities inferred from stepped leader measurements.  The corona sheath timescale depends on the time needed for charge to leave the channel, and is chosen to be $0.5$ $\mu$s as a compromise between the rapid motion of charge away from the channel out to 0.1 m distances on a timescale of 0.1 $\mu$s (i.e. a streamer speed of $\sim 1$ mm/ns \citep{briels2008pan}) and slower migration out to larger radii.  The initial charge motion outward has a stronger effect on electric fields than motion to larger radii, so the corona timescale is shorter rather than longer.  The heat capacity of the channel of $2$ J/K/m with a channel radius of 3 mm corresponds to a heat capacity of $\sim 50$~J/K/g for air at or below atmospheric pressure, consistent with experimental estimates \citep{D'Angola2007}.

The parameters specific to the simulation, the time-step size and channel segment length, are chosen based on the desired resolution.  We seek to resolve processes on shorter than 10 $\mu$s scales, requiring time-steps shorter than 0.1 $\mu$s, so here we use 35~ns.  35~ns time-step interval, given our Von Neumann stability analysis, requires segments at most $\sim10$~m long, so here we use 5~m.   We tested a variety of other spatial and temporal resolutions, and the values given above ensure good convergence, retain resolution, and are quite stable.

The simulation technique described above is applicable to many problems in lightning physics.  The geometry of the channel is unconstrained, so branched channels with arbitrary shape and connectivity are allowed.  The time-step is flexible, allowing efficient simulations both of large-scale channel development and short-duration charge motions.  Inclusion of stochastic channel extension motivated by fractal geometry allows for simulations of the full lightning discharge.  The radiative terms in the EFIE allow for prediction of electromagnetic emissions from lightning channels.  Those advanced features aside, however, we start simple in this paper.

\section{Simulation results}
\label{sect:results}

As the focus of this paper is preliminary breakdown pulses, consider simulations of single steps.  The simplest configuration that captures this phenomenon is a single isolated straight channel, with an uncharged non-conductive ``step'' portion at one end as shown in Figure~\ref{fig:seggeom}b.  This initial channel is 1~km long, unbranched, and straight, with its bottom end at 5~km altitude.  Since the discharge is so far from the ground, the effects of the charges induced in the ground by slowly-varying thunderstorm charges can be included as part of the applied field as described above.

For simulation of a step, since inter-step intervals are relatively long, we assume that the charge distribution along the channel has reached equilibrium prior to the step.  We simulate this by initially allowing charge and current to flow on the main conductive channel for a time much longer than that needed for equilibrium to be reached without allowing the step to evolve in any way.  Once the main channel has reached equilibrium, its resistance is set to 48~$\Omega$/m to represent an existing active channel and the resistance of the step portion of the channel is set to an initial non-infinite but very large value ($8 \times 10^4$ $\Omega$/m, corresponding to an initial temperature of 1400~K in our conductivity scheme).  This artificial heating of the new step to the point where it can further heat itself by current flow hides the details of the near-channel physics that somehow leads to step-wise channel extension.  The physics of this process is not well-known, though it presumably includes the effects of electric field-induced ionization, photoionization, corona, and streamer behavior ahead of the existing channel.  Regardless of the details, once our crude approximation of the initial temperature increase has been applied, the entire system is allowed to evolve freely.  The applied field and the field from charges accumulated on the main channel then drive small currents on the step which gradually heat the step until the positive feedback from rising temperatures and increasing conductivity results in a rapid current pulse from the main channel onto the newly-active step.

Given the large charge accumulation at and near the end of the main channel, the electric field is strongest at points on the step closest to the main channel.  Thus, for our geometry, the current and resulting heating are strongest close to the main channel.  If allowed to evolve without any additional requirements, the step heats starting closest to the main channel in a process akin to a dart leader.  This is logical, but contrary to observations of stepped leaders that, as discussed above, seem to involve space stems and leap forward at velocities faster than those of dart leaders \citep[see speed and duration estimates in][sections 4.4.6 and 4.7.2]{Rakov2007}.  Creating a space stem artificially entails careful tuning of the initial resistance over the newly-evolving step.  As there is no clear justification for why or how we should accomplish this, we instead take a more blunt approach and enforce a uniform resistance per unit length of the step at each time-step.  This is done by calculating the total heating and specific heat of the step channel and calculating the average effects on resistance.  Though artificial, this smoothed resistance structure means our results are not tied to any particular ideas about pre-step channel structure.  Though the resistance is enforced to be uniform, the current and charge density evolve without any smoothing.

\begin{figure}
\includegraphics[width=7in]{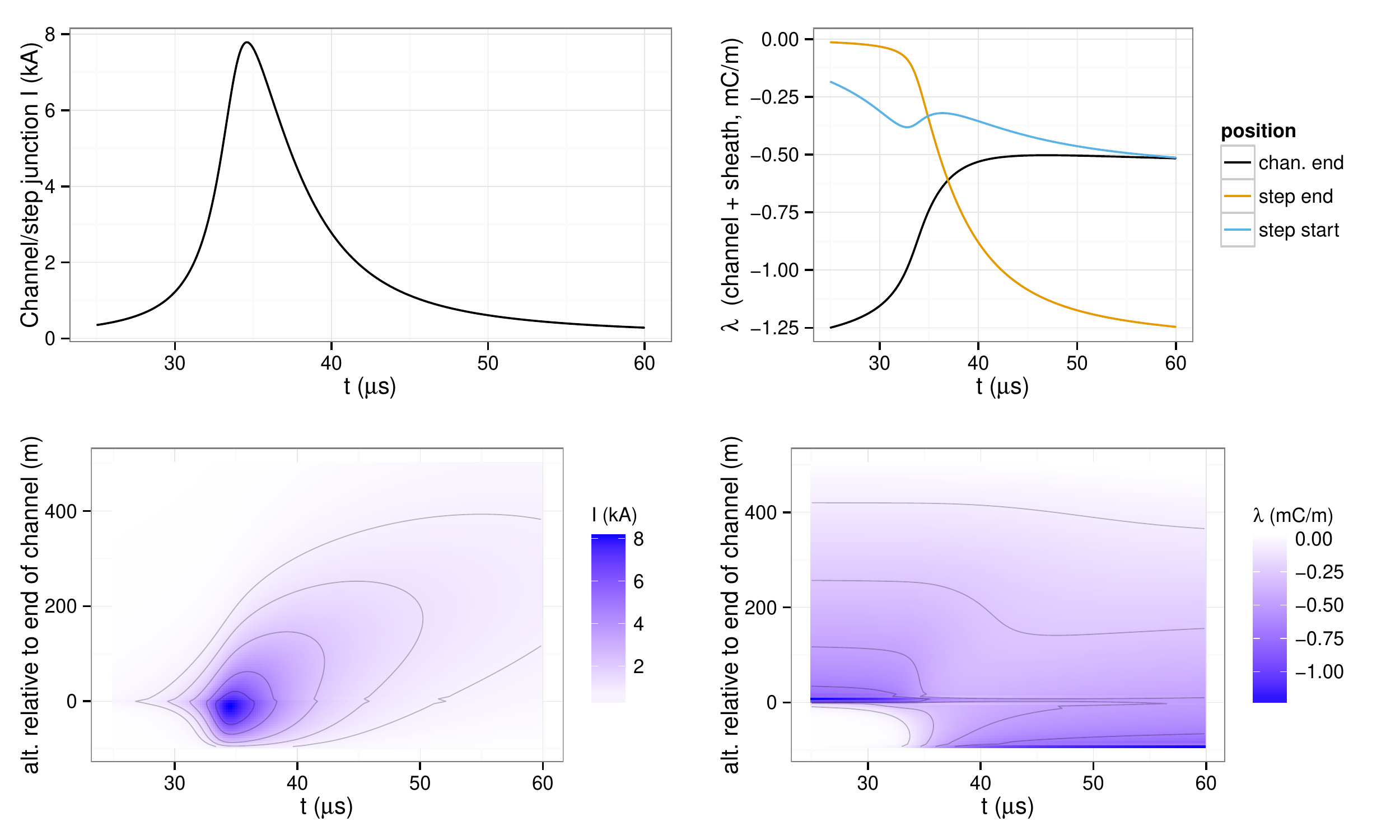}
\caption{Sample results of a 100~m step simulation.  Current (left) and charge (right) are shown vs time at various points (top) and as an image (bottom).  The current flow shown at top left is through the channel-step junction.  The charge shown at top right includes both channel and sheath charge and is plotted at three locations: the end of the old channel, the start of the step (just past the end of the channel), and the end of the step (the end of the new channel).  The images at bottom both show the evolution of the bottom 600~m of the channel, i.e. the bottom half (500~m) of the old channel with the 100~m step at the bottom.}
\label{fig:sampleresults}
\end{figure}

Sample simulation results are shown in Figure~\ref{fig:sampleresults}.  The first feature of note is that there are no high-speed propagating features since this process is dominated by resistance and heating that happen on a 10~$\mu$s timescale, much longer than the speed of light propagation time over the region shown.  The contours on the current plot (Figure~\ref{fig:sampleresults} bottom left) do suggest some propagating feature moving along the channel, but this is basically a diffusion of charge, not a return-stroke like current pulse.  Second, since current and charge evolve on the new step without any smoothing (only temperature and conductivity are smoothed), the current flow on the step first becomes significant near the end of the channel and is highest there.  This also explains the kink in the ``step start'' charge density (Figure~\ref{fig:sampleresults} top right): initially this charge density increases before the step effectively turns on, but at around 35~$\mu$s when the current is large, much of it carries charge away from the beginning of the step, decreasing the magnitude of the local charge density before charge flow from elsewhere on the channel catches up.  Third, the charge density associated with the end of the channel lingers for a relatively long time.  This is expected, even given the relatively short $\tau_\mathrm{CS}$ used in the simulation, since $\tau_\mathrm{CS}$ is the timescale for charge transfer \emph{to} the corona sheath as driven by the focused charges on a given channel segment channel, while charge transfer \emph{from} the corona sheath is driven by the charge on the sheath itself, which exerts its effect only on portions of the channel away from the segment in question and thus is a relatively slow indirect effect.  As the extension comes to equilibrium, the linear charge density is consistent with the $\lesssim 1$~mC/m inferred from measurements near lightning leaders \citep[e.g.][]{Winn2011}.

For comparison with data, the resulting currents and charges are used to calculate the electromagnetic fields observed by hypothetical receivers positioned on the ground at various positions near the channel as shown in Figure~\ref{fig:seggeom}b.  While the ground is sufficiently far below the channel to neglect the effects of charges induced in the ground by lightning on the lightning itself, this is not true for observers on the ground.  Here we simply treat the ground as a perfect conductor and apply the method of images, a reasonable first approach since ground conductivity for earth of $\sim 10^{-3}$ S/m \citep{itu1992} gives a relaxation time of $\sim 10$ ns, shorter than the processes we consider here.  The electromagnetic wave radiated by the image superposed with the electromagnetic wave radiated by the channel itself result in valid perfect-conductor boundary conditions at ground level, simply cancelling out the horizontal electric fields at the location of the receiver and doubling the vertical electric field.  This can easily be shown by consideration of the geometry of the image charges and currents and working out the vector geometry in the EFIE.  Since our applied electric field is only intended as a driver of processes on the channel itself, it does not capture screening charge layers on the cloud or local to the receiver, so we only consider changes in electric field due to charge motions during the step, not the overall field.

\begin{figure}
\includegraphics[width=7in]{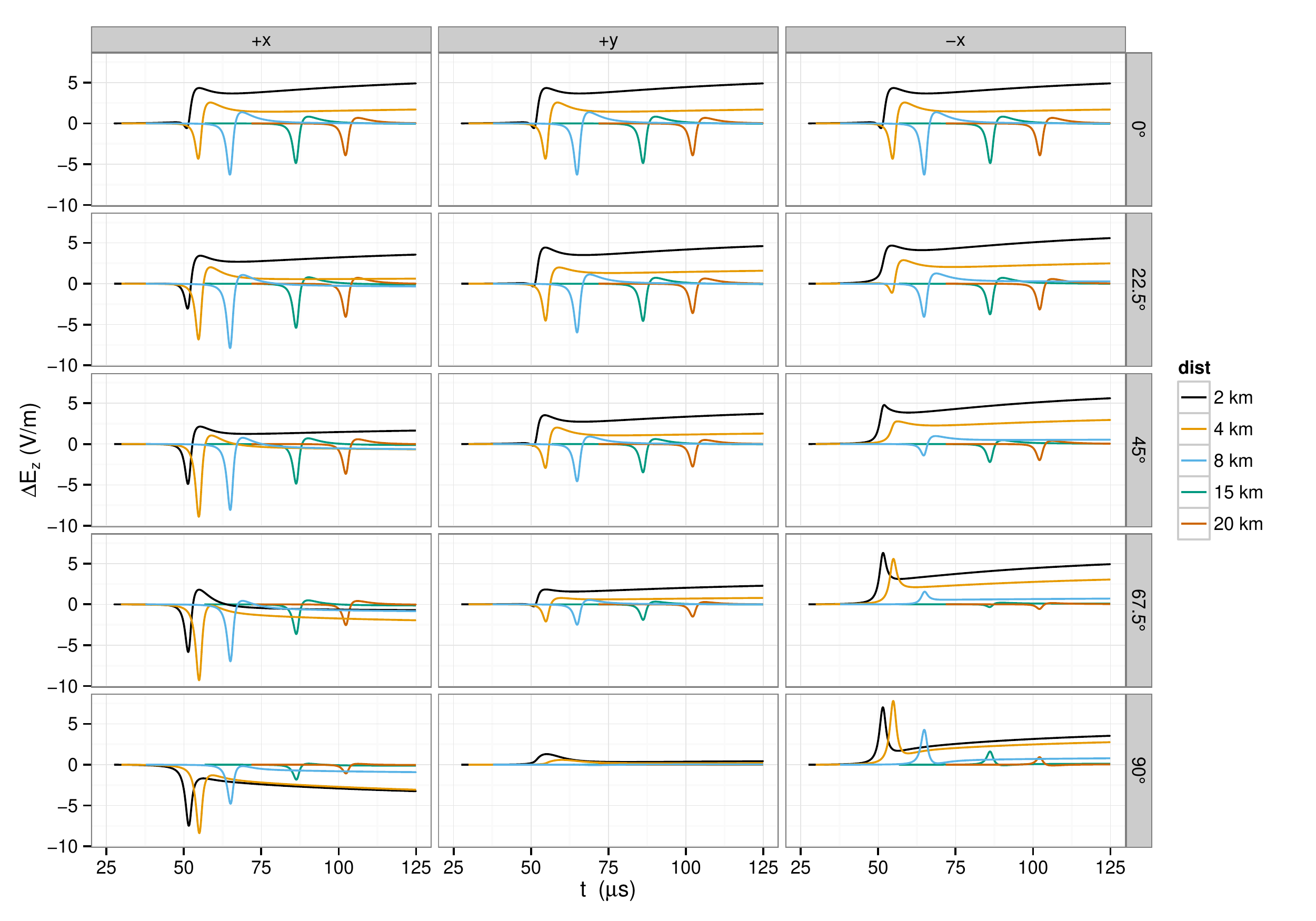}
\caption{Sample electric field change recordings at a variety of positions relative to the discharge for a variety of discharge orientations.  The geometry is as shown in Figure~\ref{fig:seggeom}b, but with the top of channel tilted away from the vertical $z$ axis toward the $+x$ axis, pivoting about the junction between channel and step.  Each row of plots in the figure corresponds to a different orientation of channel, with the angle marked at right in degrees deviation from vertical as the channel tilts.  Each column in the plot shows predicted observations at ground level at locations displaced from the sub-step point in a given direction ($+x$, $+y$, or $-x$).  Each curve in the plot shows the predicted field change for a receiver at a distance indicated by the color of the curve.  For example, the black curve in the upper left plot shows the observations predicted at a point 2~km in the $+x$ direction from the sub-step point with a vertical channel.}
\label{fig:sampleresultsAngle}
\end{figure}

Samples of such electric field change records are show in Figure~\ref{fig:sampleresultsAngle}.  Each panel shows the predicted observations at locations displaced a variety of distances from the sub-step point in a given direction for a discharge oriented at a given angle.  The simulations shown use a 100~m step on the bottom of a 1~km channel in a 100~kV/m applied electric field directed along the channel.  The signals received depend strongly on the direction of the channel and on the location of the observation point, but some trends are evident.  In all panels, observers far from the sub-discharge point observer smaller static field changes, while the radiated pulse (amplitude $\propto 1/R$) remains evident out to long distances.  For observers displaced in the $+x$ direction (the left column of plots), the upright channels result in a net transport of negative charge toward the observer (and thus producing an upward-directed static electric field change), while steeply tilted channels move negative charge on average away and positive charge toward the observer (and thus producing a downward-directed static electric field change).  No such sign change is present in observation locations displaced along the $-x$ axis (right column), since regardless of angle, the step results in a net transport of negative charge closer to the observer, though such observers see a sign change in the \emph{radiated} signals as the channel tilts and thus hits the observer with signals radiated in different directions relative to the channel orientation.  The observations displaced along the $+y$ axis (the middle column, observers for which the channel is neither angled toward nor away from the observer) fall generally in between the corresponding observations for observers displaced along the $+x$ and $-x$ axes.  One exception to this general trend is for horizontal channels, where observers along the $+y$ axis are predicted to detect very small DC field changes due to a motion of negative charge on average toward the observer, together with a small transient non-radiated pulse contributed by the $\partial\rho/\partial t$ term in the EFIE.

Clearly, even a single stepwise channel extension event can produce a wide variety of static and radiated electric field changes.  We hope at the very least that these results will be useful in qualitative interpretation of data, and with known geometry constraints or the plausible assumption of a vertical channel, such simulations can illuminate quantitative connections with individual pulses.

\subsection{Parameter dependence}
In order to complete our discussion of simulation results, we examine the dependence of the simulation results on the step properties and physical parameters of the model.  Throughout our discussion we have given plausibility arguments and citations for parameter values, but these parameters are at least slightly tunable since the values are either not known precisely or appear only as ``effective'' values.  Tunable parameters remove some of the predictive power of such a model, but tuning the parameters to match observations provides information about the allowable effective values of the parameters, making a connection between observations and more fundamental processes.

\begin{figure}
\includegraphics[width=3.0in]{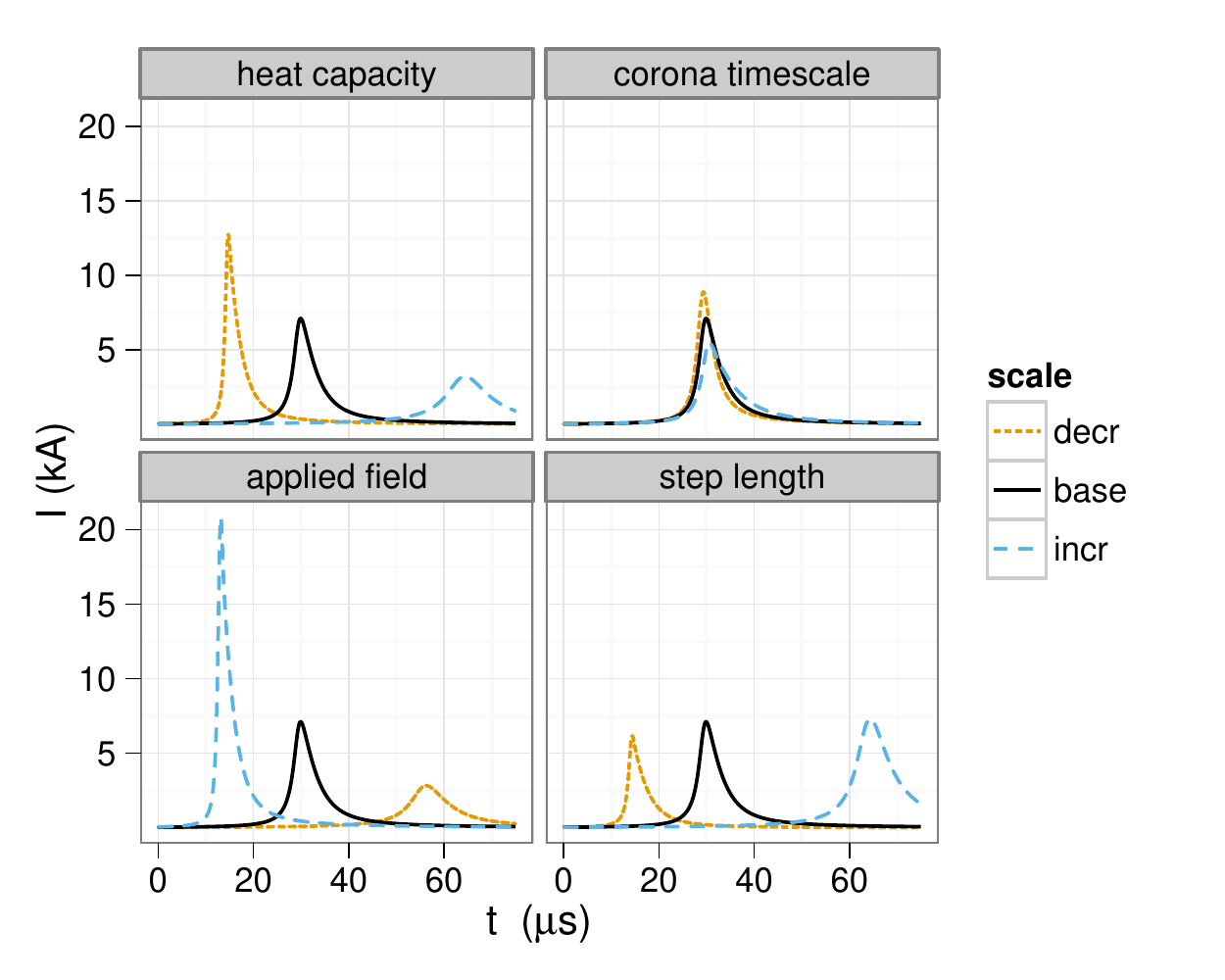}
\caption{Parameter dependence of channel current at the base of the step, showing the effects on the current if the major parameters are increased or decreased as labeled (increase or decrease is by a factor of 2 except for applied field, which is increased and decreased by 25\%).}
\label{fig:paramsI}
\end{figure}

\begin{figure}
\includegraphics[width=3.0in]{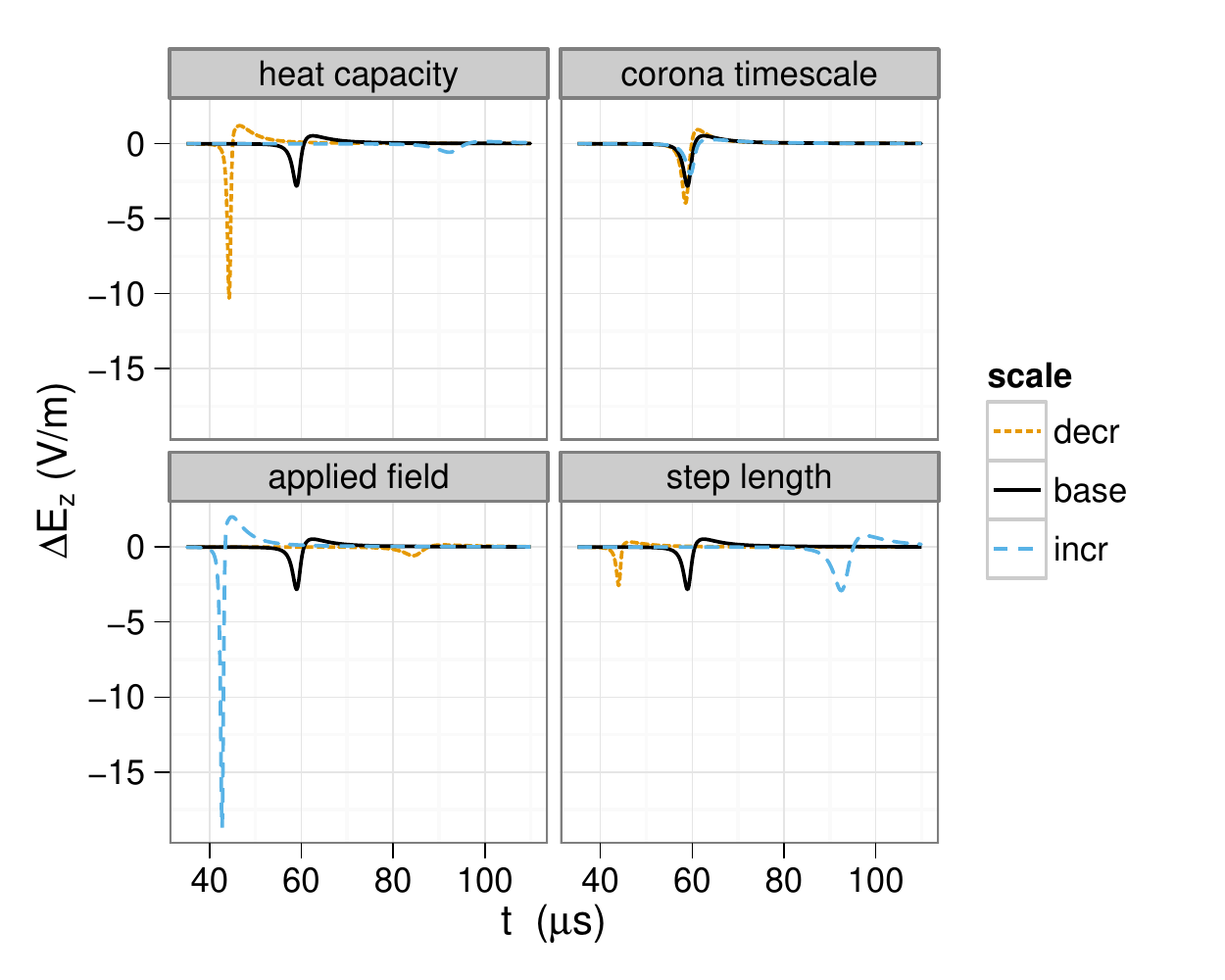}
\caption{Parameter dependence of received electric field showing the effects on the signal if the major parameters are increased or decreased as labeled (increase or decrease is by a factor of 2 except for applied field, which is increased and decreased by 25\%).}
\label{fig:paramsEZ}
\end{figure}

As a way to study the parameter dependence, consider the signals detected by a hypothetical observer a moderate distance from the sub-step point.  Provided the channel is not too steeply angled, all observers more than a few km from the sub-step point agree on the shape of the radiated electric field pulse, so this is a useful diagnostic that is somewhat less dependent on the details of the geometry.  Sample simulated current and radiated electric field waveforms are shown together with the effects of the most important parameters in Figures \ref{fig:paramsI} and \ref{fig:paramsEZ}.  During the step, the step channel heats such that its resistance decreases to 30 -- 70 $\Omega/m$, consistent with $R_l$ of the main channel and thus with growth of the main channel.  Overall, the electromagnetic signals produced (Figure \ref{fig:paramsEZ}) have 3 main identifiable features: amplitude, duration, and the relative height of negative and positive excursions (``asymmetry'').  For simulations of single lightning leader steps, extensive numerical exploration shows that the most important parameters are the specific heat of the channel, the timescale over which charge migrates to the corona sheath, the step length, and the applied electric field strength which interacts with the channel length.  As the specific heat of the channel increases, the duration of the pulse increases, its amplitude decreases, and the pulse becomes more asymmetric.  As the corona timescale increases, the pulse amplitude decreases and the pulse becomes more asymmetric, leaving the pulse duration unaffected.  Longer steps take longer to heat over their entire length and thus radiate longer duration pulses with similar amplitudes and increased symmetry.  Finally, increasing the applied electric field strength is similar to decreasing the channel heat capacity: pulse duration decreases, amplitude increases, and asymmetry decreases.  Unfortunately, applied electric field affects results similarly to the length of the pre-existing channel: longer channels lead to greater intensification of the electric field in the region of the step, and the equivalence between non-channel applied electric field and channel electric field makes long channels produce steps essentially identical to shorter channels in stronger applied electric fields.

Physically, these pulse features and their parameter dependence shed light on the physical origin of the pulse features.  First, note that the radiated electromagnetic wave comes largely from the $\partial J/\partial t$ term of the EFIE.  Strong radiated electric fields thus correspond to rapidly-changing currents.  The initial negative excursion comes from a rapid increase in upward current flowing onto the new step (the ``turn-on'' phase), while the smaller positive excursion that immediately follows comes from the decrease in current as the step gradually fills with change (the ``turn-off'' phase).   The risetime of the pulse is directly connected to the rise-time of the current, and thus to heating of the channel.  The duration of the initial negative excursion is simply the time required for the current to reach its maximum value as the channel heats, which is determined both by the energy necessary to heat the channel (heat capacity) and the amount of energy available (applied field).  The maximum current is determined by the length of the step and also by the charge accumulation necessary to counteract the applied field and thus reflects the formation of the corona sheath: rapid sheath formation draws more charge away from the channel and thus leads to higher currents since charge on the corona sheath is less able to counteract an applied field on the channel than charge on the channel itself.  Once the current reaches its maximum value, it decreases on a timescale determined by the formation of the corona sheath.  With this physical framework in mind, the simulation results can be compared to observation.

\section{Comparison to observation}

\label{sect:comparison}
These simulation results can easily be compared to data collected with the Huntsville Alabama Marx Meter Array (HAMMA).  HAMMA is a network of electric field change meters (Marx meters) located in the area surrounding the University of Alabama Huntsville.  The electric field change meters have 100 ms time constant and are sampled at 1 MHz with GPS time synchronization.  These meters provide high-resolution, high-dynamic-range measurements of electric field changes associated with both slow and fast processes in lightning discharge.  GPS time accuracy allows the location of fast processes to be determined by time of arrival fitting.  The Alabama Lightning Mapping Array (LMA) \citep{Goodman2005}, a VHF time of arrival lightning mapper, covers the same area.  

\begin{figure}
\includegraphics[width=6.0in]{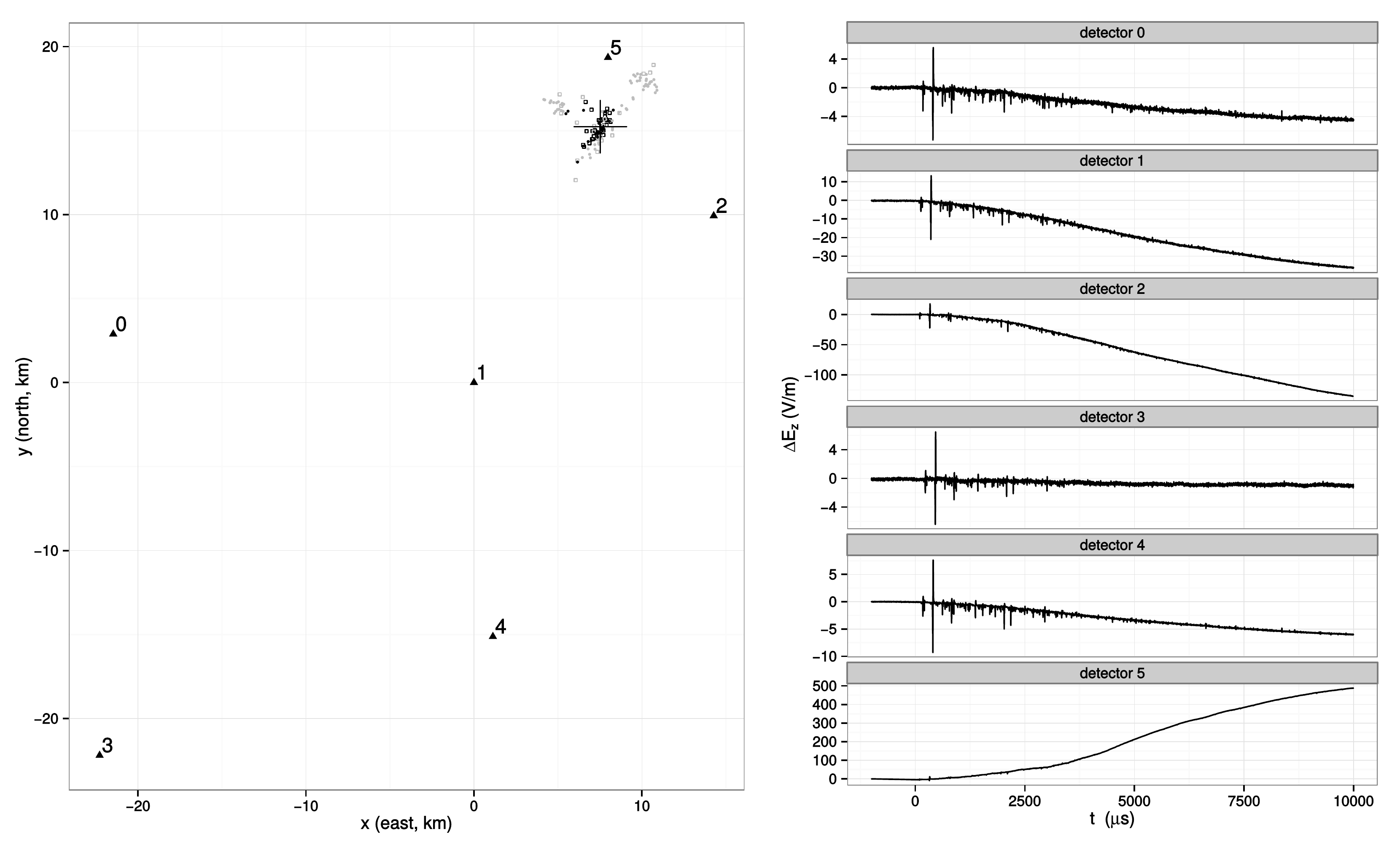}
\caption{A map of the discharge according to LMA and HAMMA time of arrival (left) and an overview of the HAMMA data for the preliminary breakdown phase (right).  At left, LMA points during and after the preliminary breakdown period shown in the data overview are shown as black and grey circles, respectively, while HAMMA time of arrival locations during and after the preliminary breakdown period are shown as black and grey open squares, respectively.  The median HAMMA time of arrival position for the preliminary breakdown period is shown as a large $+$, and has an altitude of $\sim 5$~km.}
\label{fig:dataoverview}
\end{figure}

In this paper, we focus on fast pulses measured during the initial growth of a lightning discharge on October 26, 2010 at 19:04:59 UT.  This lightning discharge lasted more than 100 ms, and included multiple K-changes and return strokes, but here we focus on the preliminary breakdown pulses during the growth of the channel just after initiation and prior to the first K-change.  A map of the discharge and the preliminary breakdown period in question is shown in Figure~\ref{fig:dataoverview}.  HAMMA detector 5 is 4.4~km from the sub-median point, and sees a positive $\Delta E_z$ due to negative charge motion toward the detector, while the other detectors are far enough away that they see a net negative $\Delta E_z$ that can be understood in the context of the curvature of electric field lines of a dipole.  For the rest of this paper, we will examine three representative detectors: detector 5 (4.4~km away, very close to the discharge), detector 2 (8.4~km away, moderate distance), and detector 4 (30.8~km, relatively large distance).

\begin{figure}
\includegraphics[width=6.0in]{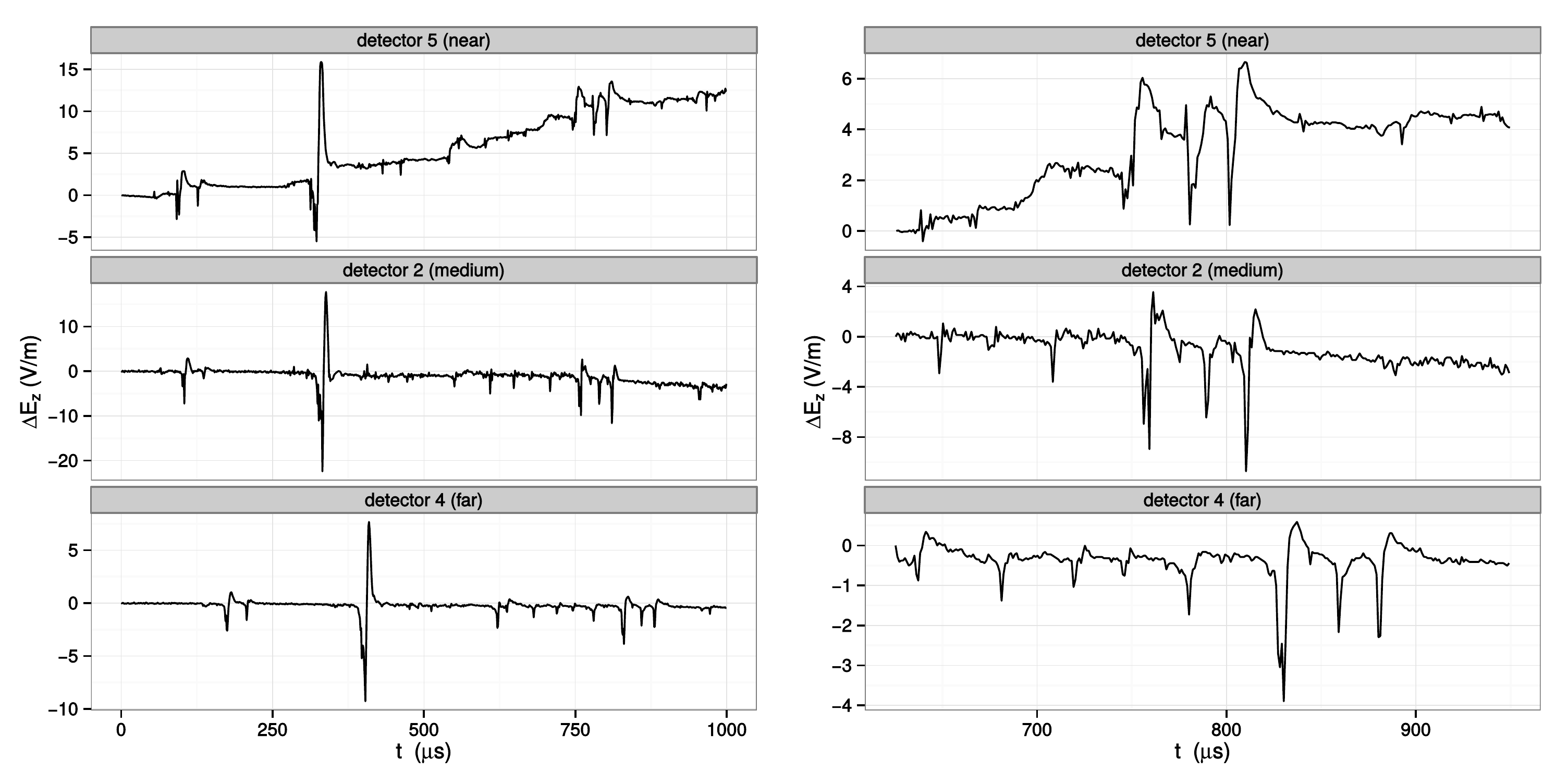}
\caption{Zoomed views of the data from Figure~\ref{fig:dataoverview}, showing the first millisecond of data (left) and a further zoom into the second half of that first millisecond.  Data from detectors 2, 4, and 5 are shown as representative.}
\label{fig:datazoom}
\end{figure}

The preliminary breakdown pulses considered in this paper are too short to be resolved on the relatively large timescale of Figure~\ref{fig:dataoverview}.  Focusing on the first few milliseconds as shown in Figure~\ref{fig:datazoom}, pulses with a variety of features can be seen.  Most pulses in detectors 2 and 4 show the same features as those in the more distant detectors in Figure~\ref{fig:sampleresultsAngle}: a relatively short intense downward excursion followed by a relatively weak and long upward excursion.  Many of the pulses in detector 5 also show this pattern, but include a more clear stepwise increase in upward electric field, a feature present in our simulation results in Figure~\ref{fig:sampleresultsAngle} for relatively nearby detectors.

Focusing on the group of three pulses at 740--810~$\mu$s in detector 5, the first pulse shows a relatively small downward excursion associated with a relatively large DC change, while the last shows a relatively large downward excursion with very little DC change.  Comparison to our simulation results suggest the first pulse was associated with channel extension directed somewhat but not directly toward detector 5 (see the $-x$, $22.5^\circ$,  4~km curve in Figure~\ref{fig:sampleresultsAngle}), while the third pulse was associated with extension directed more perpendicular to the line of sight from detector to channel (see the $+x$, $22.5^\circ$, 4~km curve).  Detector 4, on the opposite side of the discharge as detector 5, is well-placed to test this hypothesis.  If the first pulse of the trio was toward detector 5 (matching $-x$ curves), it should have been away from detector 4 (matching $+x$ curves), while if the third pulse of the trio was away from detector 5 (matching $+x$ curves), it should have been toward detector 4 (matching $-x$ curves).  Unfortunately, the only visible difference between the $+x$ (away from, hypothetical first pulse) and $-x$ (toward, hypothetical third pulse) curves for detector 4 is amplitude, with the toward ($-x$) curve having a slightly lower amplitude.  The third pulse in detector 4 does indeed have a slightly lower amplitude than the first, consistent with the predictions of the model, but the relative amplitude of the simulation results comes from a single simulation, while in the data we are comparing two distinct pulses.  Detector 2, however, can address this uncertainty; located approximately perpendicular to the line connecting detector 5, the lightning channel, and detector 4, the symmetry of the situation suggests that whether channel extension is directed toward detector 5 or toward detector 4 should not affect the pulse observed by detector 2, so detector 2 can be used to judge the relative amplitude of pulses as emitted by the channel.  Detector 2 sees approximately equal amplitudes, which indicates that the channel extension events responsible for the first and third pulses are of approximately equal intensity.  This lends support to the comparison between a single simulation and two pulses seen in detector 4 as described above, and suggests that the amplitude difference between the first and third pulses seen in detectors 3 and 4 can be attributed to the different directions of channel extension relative to detector location.  Our interpretation of channel directions and pulse intensities as seen by detectors 4 and 5 as motivated by the pulse shapes seen by detector 5 and supported by detector 2 is thus at least qualitatively self-consistent.

\begin{figure}
\includegraphics[width=6in]{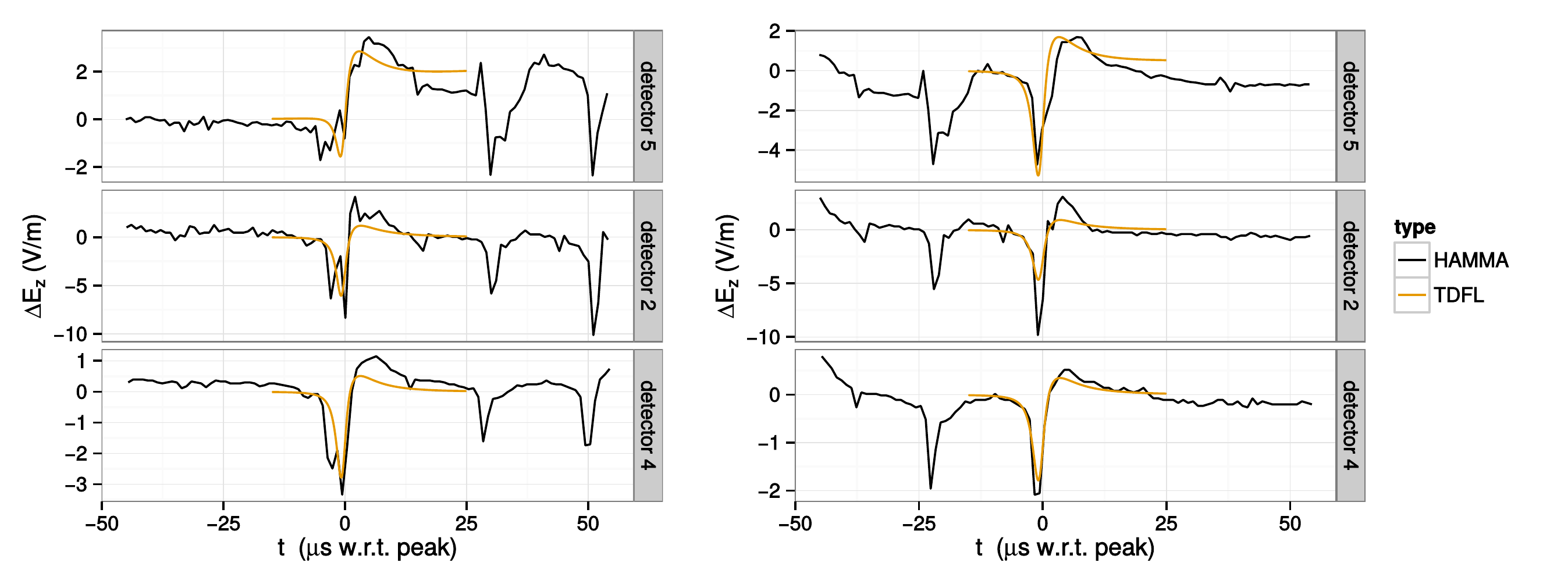}
\caption{Quantitative comparison between simulation results and observations for the pulses discussed geometrically in the text (i.e. those at $\sim 800$ $\mu$s in Figure~\ref{fig:datazoom}), showing at left and right the first and third pulses in the group of three, respectively.}
\label{fig:simAndData}
\end{figure}

Quantitative consistency requires direct comparison and manual iterative adjustment of simulation initial conditions.  The results of such a process for the first and third pulses discussed above are shown in Figure~\ref{fig:simAndData}.  The geometry of the simulations is exactly as in Figures~\ref{fig:seggeom}b and \ref{fig:sampleresultsAngle}, with the channel tip placed at 5~km altitude and positioned relative to the detectors as suggested by the median HAMMA source as shown in Figure~\ref{fig:dataoverview}.  We slightly adjust the channel direction to attempt to fit the observations, changing the angle from $22.5^\circ$ to $20^\circ$ to better emphasize the initial downward excursion at detector 5.  The quantitative agreement is much better for the third pulse than the for the first.  This is unsurprising given the fine structure evident in the first pulse of the three; the first pulse is likely due to a more complex process than a simple single step extension, perhaps a superposition of two overlapping extension events as suggested by the two negative excursions visible in detector 2 data.  The third pulse as seen in detector 2 is stronger than expected based on the simulation.  This suggests the directionality of the channel extension is not as simple as described above.  Adjusting the directionality such that detector 2 receives more of the radiated electric field can improve the match, as does moving the simulation channel closer to detector 2, but an automated fit would be required to improve the results significantly, the time required to run such simulations makes this difficult, and the match between a simple simulation and a complicated lightning channel is not expected to be perfect.

The comparison in Figure~\ref{fig:simAndData} does point out some detailed qualitative deviations between simulation results and observations.  First, the simulated DC field change that develops by around 25 $\mu$s after the pulse often disagrees with the observations.  This suggests either the amount of charge transferred on the simulated step is wrong or there is some other charge transfer occurring elsewhere on the channel that confounds the data.  Second, the positive excursion peaks later in the data than in the simulation.  For radiated pulses, this indicates that the current begins decreasing more quickly in the simulation than in reality.  The fact that the positive excursion is often higher in the data than in the simulation indicates that the maximum rate of decrease of current is higher in reality.  Relative to simulation, therefore, in nature, the current that flows onto a new portion of channel increases in roughly the same way, but remains near its peak for longer before turning off more quickly.  Such deviations are perhaps unsurprising given our enforced ignorance of the geometry of the heating of the step and our approximate treatment of the corona sheath, but broad qualitative agreement is promising, especially given that the data do not come from a straight channel that comes to equilibrium before extending in a single isolated step, but from active development throughout a dynamic branched channel.

\section{Summary and future work}
\label{sect:discussion}

This paper describes a simulation technique that captures the details of charge and current flow on an evolving lightning channel.  The simulation includes approximate treatments of channel resistance evolution due to heating and the migration of charge outward from the channel to the corona sheath.  Inclusion of these processes leads to a model capable of reproducing the detailed features of preliminary breakdown pulses as shown in Figure~\ref{fig:simAndData}, lending support to the interpretation of such pulses as from stepwise extensions of an existing channel.

Much work remains to be done, however, as seen both in the deviations between simulation and data and in the fact that we only consider a small portion of the overall evolution of the channel.  In this, the mismatch between simulation and observation is encouraging; such mismatch means the results of the simulation are sensitive to the details of the processes at work in a preliminary breakdown pulse, so further study can shed light on such details.  For example, the framework described here can be extended to include more detailed treatments of the plasma physics of the channel (see for example \citet{Liang2014}), and the resistance of the new step channel, here forced to be uniform, can be allowed to vary with better initial conditions, perhaps approximating a space leader process.  The channel extension process can be simulated further by including more steps and variation of step properties with altitude.  On longer timescales, the simulated channel can be allowed to  extend and branch stochastically by implementing results from fractal lightning (e.g. \citet{Niemeyer1984,Riousset2007}, justifying our name for the technique as time-domain fractal lightning modeling), and preliminary results show excellent qualitative agreement with longer timescales of channel evolution (e.g. reproducing K-changes), to be described in a subsequent paper.  Finally, though the simulation reproduces many features of lightning electric fields, there are still features that are difficult to explain, like the unusually large and symmetric pulse in Figure~\ref{fig:datazoom} near $375$ $\mu$s.  Such a feature must represent a current pulse that turns on rapidly and turns off just as rapidly, suggesting that the effect of processes like the formation of the corona sheath are not as important for such pulses.  Such speculation can easily be tested, and the future work described above is ongoing.  It is our hope that such full electrodynamic simulations, motivated by physics, can help bridge the gap between plasma physics and lightning observation, helping both to constrain our understanding of the physics of the lightning channel and to interpret lightning observations.

\begin{acknowledgments}
The HAMMA data used in this paper are available from the authors on request.

The authors are very grateful for useful discussions with Thomas Gjesteland, Nikolai {\O}stgaard, and Forrest Foust, and especially Justin Barhite for assistance and sanity-checking the von Neumann stability analysis.  Referee comments on a previous draft were also very helpful.  This work was supported by Norwegian Research Council grant 197638/V30, US National Science Foundation grant ATM-0836326, DARPA NIMBUS grant HR0011-10-1-0058, and NASA grant NNM05AA22A. 
\end{acknowledgments}


\end{article}

\end{document}